\documentstyle[aas2pp4,11pt]{article} 
\begin{document}

\title{Quasi-periodic oscillations discovered in the new
X-ray pulsar XTE~J1858+034}

\author{B. Paul \and A. R. Rao}
\affil{Tata Institute of Fundamental Research, Homi Bhabha Road, Mumbai 400 005, India}

\begin{abstract}

We report the discovery of low frequency quasi-periodic oscillations
centered at 0.11 Hz in the newly discovered 221 s X-ray pulsar XTE J1858+034.
Among about 30 known transient X-ray pulsars this is the sixth source in
which QPOs have been observed. If the QPOs are produced because of
inhomogeneities in the accretion disk at the magnetospheric boundary, the low
frequency of the QPOs indicate a large magnetosphere for this pulsar. Both
the Keplerian frequency model and the beat frequency model are applicable
for production of QPOs in this source. The QPOs and regular pulsations are
found to be stronger at higher energy which favours the beat frequency model.
The magnetic field of the pulsar is calculated as a function of its distance.
The energy spectrum is found to be very hard, consisting of two components,
a cut-off power law and an iron fluorescence line.

\end{abstract}

\section{Introduction}

Quasi periodic oscillations (QPOs) observed in X-ray binaries are generally
thought to be related to the rotation of the inner accretion disk.
When the accretion disk can reach very close to the compact
object, like in the case of black hole candidates and low magnetic field
neutron star sources, the rotation of the inhomogeneities or hot blobs of
material in the inner disk are reflected in the light curve as QPOs. In X-ray
pulsars, however, the disk is interrupted at a large distance by the strong
magnetic field of the neutron star, and the inner transition zone of the disk,
which is at a large distance from the neutron star, does not emit in X-rays.
Hence strong QPOs are believed to be rare in X-ray pulsars.

The hard X-ray transient XTE J1858+034 was discovered with the RXTE All Sky
Monitor (ASM) in 1998 February (Remillard \& Levine 1998). The spectrum was
found to be hard, similar to the spectra of X-ray pulsars. Observations were
made immediately after this with the Proportional Counter Array (PCA) of the
RXTE and regular pulsations with a period of $221.0 \pm 0.5$ s were discovered
(Takeshima et al. 1998). The pulse profile is found to be nearly sinusoidal
with a pulse fraction of $\sim 25\%$. From the transient nature of this source and
pulsations they suggested that this is a Be-X-ray binary. The position of the
X-ray source was refined by scanning the sky around the source with the PCA
(Marshall et al. 1998).

From the XTE target of opportunity (TOO) public archival data of the
observations of XTE J1858+034, made in 1998 February 20 and 24, we have
discovered the presence of low frequency QPOs. We also have obtained the pulse
profile of this source in two energy bands and the energy spectrum in one
of the observations. In the following sections we describe the archival data
that has been used, the analysis and results and discuss some implications
of the detection of QPOs in this source.

\section{Data}
 
We have analysed four RXTE/PCA observations of XTE J1858+034 made in the
high state of the source. The raw data was obtained from the XTE Science
Operations Facility (SOF) public data archive. We generated light curves
from the raw data in two energy bands 1.3$-$7 keV and $>$ 7 keV with
1 s time resolution. Details of the data used for this analysis, with
start and end time of the data stretches, average PCA background
subtracted count rates etc. are given in Table 1. The light curves obtained
from these observations are shown in Fig. 1 with 5 s time bin.
Information about the PCA detectors and the RXTE can be found in Jahoda
et al. (1996).

{\small
\begin{table}[h]
\caption{The observation log}
\begin{flushleft}
\begin{tabular}{lcccl}
\hline
Obs.       &Observation duration         &Useful   &Count      &\\
           &     (UT)                    &exposure &rate       &\\
           &                             &time (s) &($s^{-1}$) &\\
\hline
A          &1998 Feb. 20  21:52 to 22:08 & 934     &138        &\\
B          &1998 Feb. 24  10:53 to 11:24 & 1918    &167        &\\
C          &1998 Feb. 24  15:31 to 16:10 & 2318    &144        &\\
D          &1998 Feb. 24  17:07 to 17:46 & 1702    &148        &\\
\hline
\end{tabular}
\end{flushleft}
\end{table}
}

\section{Results }

\subsection{Periodic pulsations}

Pulsations with 221 s period as reported by Takeshima et al. (1998) are clearly
seen in the light curves (Fig. 1). We have used all four observation
stretches and obtained the pulsation period by $\chi^2$ maximising method. The
period obtained thus is $220.7 \pm 0.1$ s. The pulsation period could not be
determined very accurately because of only about 7000 s of useful data and a
large pulse period. Barycentric corrections were not applied which is about
one order of magnitude smaller than the error in period estimation. All the
light curves were folded with this period and the resultant pulse profiles in
two energy bands are shown in the top two panels of Fig. 2 for two cycles.
The pulse profile is single peaked and nearly sinusoidal as reported earlier
by Takeshima et al. (1998). The pulse fraction (defined as the ratio of pulsed
flux to total flux) in the higher energy band ($>$ 7 keV) is 20\%, significantly
larger than that of 10\% in the lower energy band (1.3$-$7 keV). There is
indication of significant change in the spectrum with the pulse phase, as the
hardness ratio, shown in the bottom panel of Fig. 2, varies by about 15\%
during the pulsation. A detailed analysis of pulse phased spectrum is in
progress.

\subsection{Power density spectrum}

We generated power density spectrum (PDS) from the 1 s time resolution data.
The light curves were broken into segments of length 512 s and the PDS
obtained from each of these segments were averaged to produce the final
PDS as shown in Fig. 3. A broad QPO feature around 0.1 Hz is very prominent
in the PDS. The PDS in the frequency range of 0.006 to 0.6 Hz fits well with
a model consisting of a power-law type spectrum and a Gaussian ($\chi_r^2$ of
1.4 for 68 degrees of freedom). The value of $\chi_r^2$ is 2.3 when only a
power law is used indicating that the presence of the feature at 0.11 Hz is
very significant.
The power-law index for the best fit model is found to be -0.95 and the
Gaussian, representing the QPO feature is centered at 0.11 $\pm$ 0.01 Hz
with a width of 0.02 Hz. While fitting the PDS to this model, the region
below 0.006 Hz were excluded to avoid the power due to the regular
pulsations at 0.0045 Hz. The rms variability in the QPO feature is 6.5\%.
PDS were also generated in the two energy bands of 1.3$-$7 keV and $>$ 7 keV.
The PDS in the two energy bands are found to be identical in shape,
comprising of one power law component of index -0.95 and a QPO feature.
The rms variation in the high energy band is generally higher and rms in
the QPOs is much more in the higher energy band (7.8\%) compared to the
same in the lower energy band (3.7\%). There was no detectable difference
in the QPO frequency in the four data sets.

\begin{figure}[t] 
\vskip 7cm
\includegraphics{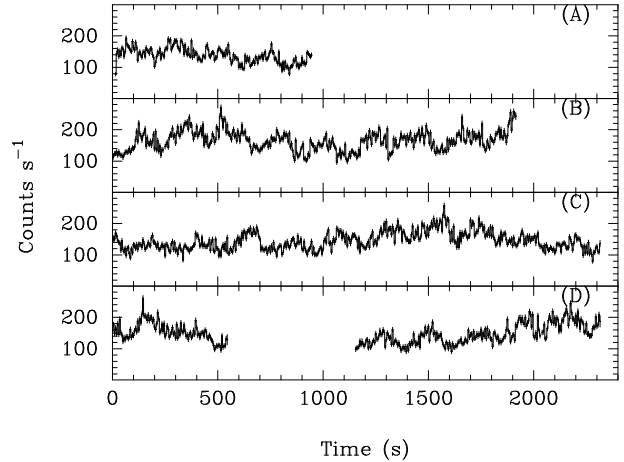} 
\caption{The background subtracted X-ray light curves of XTE J1858+034
obtained from PCA detectors in the energy range of 1.3$-$100 keV are shown
for the four observations (described in the text). The time bin is 5 s.
Modulations at the 221 s pulse period are clearly seen.}
\end{figure} 

\begin{figure}[t] 
\vskip 9cm
\includegraphics{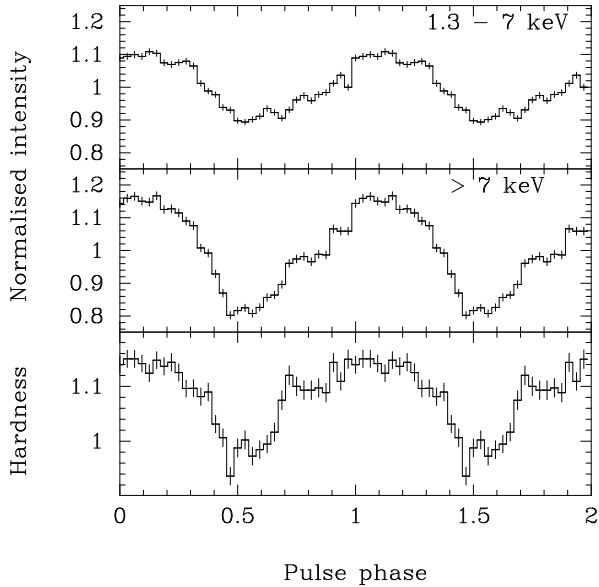} 
\caption{The pulse profiles of XTE J1858+034 folded at a period of 220.7 s
are shown in two energy bands along with the hardness ratio. The profiles
are repeated for 2 cycles for clarity.}
\end{figure}

\begin{figure}[t] 
\vskip 7cm
\includegraphics{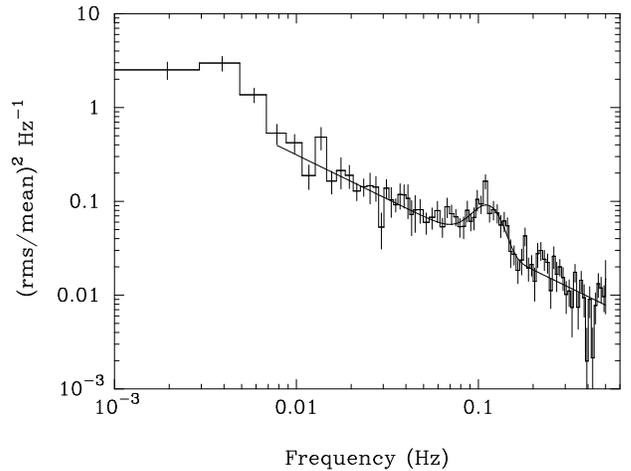} 
\caption{The power density spectrum of XTE J1858+034 generated from the
light curve over the entire energy band of the PCA. The line represents
the best fitted model in the frequency range of 0.006$-$0.6 Hz comprising
a power-law type spectrum and a Gaussian centered at the QPO frequency.}
\end{figure} 

\begin{figure}[t] 
\vskip 12.2cm
\includegraphics{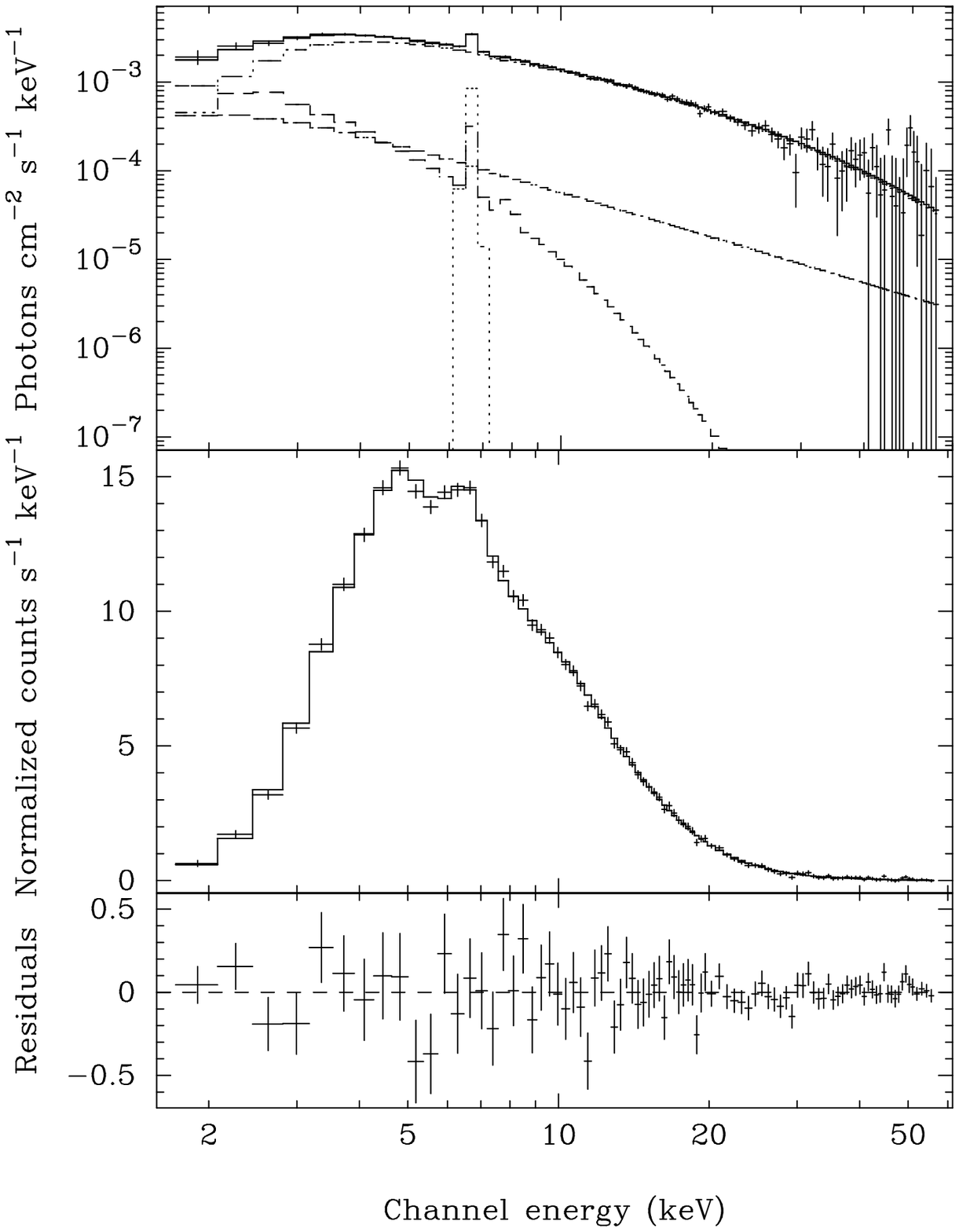} 
\caption{The X-ray spectrum of XTE J1858+034 is fitted with a cut-off power-law
of index 1 and Gaussian line at 6.6 keV of equivalent width 165 eV. The
Galactic ridge emission is modeled as a Raymond-Smith plasma of temperature
2.6 keV and a power-law of index 1.7.}
\end{figure} 

\subsection{Energy spectrum}

We have generated the count spectrum in 129 binned channels of the PCA
detectors from the observation A (see Table 1). The background was generated
using the "pcabackest" model provided by the XTE guest observer facility
(GOF). Data from all the 5 detectors were added together to produce the
spectrum. One low energy channel and channels corresponding to energy greater
than 50 keV were ignored because of low signal to noise ratio. The new pulsar
XTE J1858+034 is in the Galactic ridge (l $\sim$ 36$^\circ$.8 and b $\sim$
0$^\circ$.138). The background subtraction model that we have used takes care
of the diffuse cosmic X-ray emission and the internal background, but not the
emission from extended source like the Galactic ridge. From a detailed
observation of the Galactic ridge obtained using the PCAs (Valinia \&
Marshall, 1998) we estimate that about 10\% of the observed flux can be
accounted for by the Galactic ridge emission. We have attempted a spectral
fitting for the pulse averaged spectrum of XTE J1858+034 by explicitly taking
the Galactic ridge emission as a sum of a Raymond-Smith plasma and a power-law,
with parameters constrained to be within the range obtained for the ridge
emission in Valinia \& Marshall (1998). The residual spectrum in 1.7$-$50 keV
range is found to be very hard which can be described as a cut-off power law
with a power-law photon index close to 1, cut-off energy of 21 keV, along with
a neutral absorption
with an equivalent Hydrogen column density of 6 $\times$ 10$^{22}$ cm$^{-2}$.
A narrow emission line, which can be ascribed to atomic Iron inner shell
emission, was also found at 6.6 keV, with an equivalent width of 165 eV. Though
an acceptable value of $\chi^2$ was obtained (82 for 83 degrees of freedom),
parameters values could not be constrained due to the large number of free
parameters involved. The total incident flux in the 1.3 $-$ 100 keV band is
6.5 $\times$ 10$^{-10}$ erg cm$^{-2}$ s$^{-1}$ for the pulsar and
0.5 $\times$ 10$^{-10}$ erg cm$^{-2}$ s$^{-1}$ for the Galactic ridge emission.
The best fit spectrum with the parameters mentioned above is shown in the top
panel of Fig. 4 along with the observed spectrum deconvolved through the
detector response function. The observed spectrum and the folded model along
with the residual to the model fit are shown in the middle and the bottom
panel of the same figure.

\section{Discussion}

The transient X-ray pulsars in which QPOs have been detected, are the high
mass X-ray binaries (HMXB) EXO 2030+375 (Angelini et al. 1989), A 0535+262
(Finger et al. 1996), 4U 0115+63 (Soong \& Swank 1989) and V 0332+53
(Takeshima et al. 1994) and the LMXB GRO J1744-28 (Zhang et al. 1996; See
Finger 1998 for a review of the QPO in transient X-ray pulsars). QPOs have
also been observed in some of the persistent HMXB sources: Cen X-3 (Takeshima
et al. 1991), SMC X-1 (Angelini et al. 1991), X Persei (Takeshima 1997) and
4U 1907+09 (in'tZand et al. 1998) and the LMXB 4U 1626-67 (Shinoda et al. 1990;
Kommers et al. 1998). Both the Keplerian frequency model (in which the QPOs
are produced because some inhomogeneous structure in the Keplerian disk
attenuates the pulsar beam regularly) and the beat frequency model (in which
the material influx to the pulsar from the disk is modulated at the Keplerian
frequency) are in very good aggrement with the observations in EXO 2030+375
and A 0535+262. In 4U 0115+63, V 0332+52, Cen X-3, 4U 1626-67 and SMC X-1
however, the QPO frequency is found to be lower than the pulsation frequency
hence the Keplerian frequency model is not applicable in these sources because
if the Keplerian frequency at the magnetospheric boundary is less than the
spin frequency, centrifugal inhibition of mass accretion will take place. For
V 0332+52 the beat frequency model may also be inapplicable because the
magnetospheric boundary calculated from the QPO properties and from observed
luminosity are in disagreement in this source. In the LMXB transient pulsar
GRO J1744-28, large change in X-ray flux was found to be associated with a
very little change in the QPO frequency which ruled out both the Keplerian
and the beat frequency models for QPOs in this source (Zhang et al. 1996).
The beat frequency
model is applicable in many sources though there is no convincing evidence
of positive correlation between the QPO frequency and the X-ray luminosity
in some of them.

According to the beat frequency model, the QPOs are a
result of beat phenomena between the rotation of the innermost part of the
disk and the spin of the neutron star. The Keplerian rotation frequency
$\nu_K$ of the disk at the magnetosphere boundary, the rotation frequency
of the neutron star $\nu_S$ and the QPO frequency $\nu_{QPO}$ are related as
$\nu_{QPO}$ = $\nu_K$ - $\nu_S$. Assuming that the QPOs are produced as a
result of this phenomena, the Keplerian rotational frequency of the innermost
part of the disk is just sum of the QPO frequency and the rotation frequency
of the pulsar. For an assumed mass of 1.4 M$_\odot$, this can be related
to the magnetospheric radius r$_M$ of the X-ray pulsar.

In XTE J1858+034, we find that $\nu_{QPO}$ = 0.11 $\pm$ 0.01 Hz,
$\nu_S$ = 0.0045 Hz and the radius of the magnetospheric boundary is
calculated to be
\begin{equation}
r_M
= \left({{GM}\over { 4 \pi^2 \nu_K^2}}\right)^{1 \over 3}
= 6.5~10^8 \left({M \over {M_\odot}}\right)^{1 \over 3} {\rm cm}
\end{equation}
where M is the mass of the neutron star.

The pulse averaged X-ray flux in the 1.3-100 keV band is 6.5 10$^{-10}$ erg cm$^{-2}$
s$^{-1}$ which, for a distance of r$_{kpc}$, amounts to an X-ray luminosity
L$_X$ of 7.9 10$^{34}$ r$_{kpc}^2$ erg s$^{-1}$. For a standard accretion
disk with disk axis parallel to the magnetic field axis and dipole magnetic
field structure of the neutron star, the radius of the inner transition
zone can also be expressed as (Frank et al. 1992)

\begin{equation}
r_M = 2.9 \times 10^{8} {\left(M\over M_\odot\right)}^{1\over7} R_6^{-{2\over7}} L_{37}^{-{2\over7}} \mu_{30}^{4\over7}
\end{equation}
where, R$_6$ is the radius of the neutron star in unit of 10$^6$ cm, L$_{37}$
is X-ray luminosity in unit of 10$^{37}$ erg and $\mu_{30}$ is magnetic moment
in unit of 10$^{30}$ cm$^3$ Gauss.

Combining the above two equations, and using M = 1.4 M$_\odot$, R$_6$ = 1, 
the magnetic moment $\mu_{30}$ of the pulsar is calculated to be $\sim$
0.4 $\times$ 10$^{30}$ r$_{kpc}$, which for a neutron star radius of 10$^6$ cm,
is equivalent to a magnetic field of 0.8 $\times$ 10$^{12}$ r$_{kpc}$ Gauss. 

If origin of the QPOs in this source is the magnetospheric boundary, the
QPOs cannot arise from the modulation of X-rays emitted from the accretion
disk because for a magnetospheric radius of 3 $\times$ 10$^8$ cm the disk
temperature is rather low to emit in X-rays. The X-ray modulation at the
QPO frequency can arise either because some inhomogeneous structure in
the Keplerian disk attenuates the pulsar beam regularly at its rotation
frequency, or the material influx to the pulsar from the disk is modulated
at the Keplerian frequency. The fact that the strength of the QPO is
greater at higher energies indicates that the latter is likely to be the
case for XTE J1858+034. A detailed analysis (which is currently in
progress) of the QPO feature as a function of pulse phase and energy
will help in firmly deciding one of the two alternatives for the QPO
phenomenon.

\begin{acknowledgements}
We thank the RXTE team and M. Takeshima in particular for providing the
realtime archival data and software support. We thank an anonymous referee
and M. Finger for some valuable suggestions. This research has made use of
computer systems of the optical CCD astronomy programme of TIFR.
\end{acknowledgements}

\end{document}